\documentclass[amsmath,amssymb,aps,prb,reprint,twocolumn,superscriptaddress,longbibliography,floatfix]{revtex4-2}
\usepackage{amsmath}
\usepackage{amsfonts}
\usepackage{graphicx}
\usepackage{xcolor}
\usepackage{ulem}
\usepackage{units}  
\usepackage{float}
\usepackage{xspace}
\usepackage{subfigure}
\usepackage{caption}
\usepackage{subcaption}
\usepackage{hyperref}
\usepackage{braket}
\usepackage{url}
\usepackage{booktabs}
\usepackage{multirow}
\begin{document}
	\title{Skin-Anderson Localization Transition in Strongly Coupled Disordered Non-Hermitian Chains}
	
	\author{S Rahul}
		\affiliation{Manipal Institute of Technology Bengaluru, Manipal Academy of Higher Education, Manipal-576104, India}

	\date{\today}
\begin{abstract}
	The interplay between disorder and non-Hermitian effects gives rise to a variety of intriguing localization phenomena. While disorder tends to localize the eigenstates through Anderson localization, non-Hermitian non-reciprocity promotes the formation of skin modes by driving the eigenstates toward the system boundaries giving rise to non-Hermitian skin effect (NHSE). In this work, we investigate the interplay between these competing mechanisms in a two-leg ladder consisting of a Hatano--Nelson chain coupled to a Hermitian chain via asymmetric inter-chain hopping, with strong disorder present in both chains. We show that tuning the asymmetry of the inter-chain coupling induces successive transitions in the nature of the eigenstates, from skin localization to Anderson localization and subsequently back to skin localization. Remarkably, the non-Hermitian skin effect re-emerges even though the energy spectrum exhibits a line-gap topology, demonstrating that robust skin localization can persist beyond the conventional point-gap regime.
\end{abstract}
	
	\maketitle
	
\section{Introduction}
In recent years, non-Hermitian physics has attracted significant attention owing to its rich physical phenomena, including complex energy spectra, the NHSE~\cite{bender1998real,alvarez2018non,gong2018topological,leykam2017edge,el2018non,yokomizo2019non,PhysRevLett.122.076801,PhysRevLett.125.126402,PhysRevLett.124.086801,PhysRevB.100.054105,heiss2012physics,bergholtz2019exceptional,lee2016anomalous,PhysRevLett.125.126402,kawabata2019symmetry} and exceptional points (EPs) \cite{bergholtz2019exceptional,heiss2012physics,san2016majorana}. Among these, the NHSE has emerged as one of the most distinctive features of non-Hermitian systems, characterized by the macroscopic accumulation of eigenstates at the system boundaries under open boundary conditions~\cite{yao2018edge,lee2016anomalous}. The NHSE is closely associated with a nontrivial point-gap topology of the complex energy spectrum, leading to the non-Bloch bulk-boundary correspondence and a breakdown of the conventional Bloch description~\cite{yao2018edge,yokomizo2019non,koch2020bulk,PhysRevLett.121.026808}. This intimate connection between point-gap topology and the NHSE has fundamentally reshaped the understanding of bulk-boundary correspondence in non-Hermitian systems.\\
Considerable effort has been devoted to understanding the robustness of the NHSE in the presence of various perturbations, such as disorder~\cite{jin2025anderson}, impurities~\cite{li2021impurity,longhi2022selective}, and modified boundary conditions~\cite{edvardsson2022sensitivity}. 
Disorder plays a significant role in non-Hermitian systems, affecting entanglement and topology \cite{PhysRevB.58.8384,PhysRevLett.126.166801,PhysRevB.100.054301,PhysRevX.8.031079,PhysRevLett.122.237601,PhysRevLett.127.213601,PhysRevLett.126.090402,PhysRevB.107.144204,PhysRevLett.81.862}. 
Disorder generally promotes Anderson localization by suppressing wave propagation through multiple scattering, leading to spatially localized eigenstates. In contrast, non-reciprocal hopping inherent to non-Hermitian systems induces a directional flow of probability, causing eigenstates to accumulate exponentially at one boundary and giving rise to the NHSE. The best known model to exhibit such a phenomenon is the Hatano-Nelson model in presence of disorder \cite{PhysRevLett.77.570}. \\
In this work, we investigate the interplay between the NHSE and Anderson localization in a two-leg ladder comprising a Hatano--Nelson chain asymmetrically coupled to a Hermitian chain, with strong on-site disorder present in both chains. The asymmetric inter-chain coupling parameter $a$ controls the competition between the horizontal skin accumulation generated by the Hatano--Nelson chain and the directional diffusion between the two chains. For reciprocal inter-chain coupling ($a=0$), the two chains are strongly hybridized while the NHSE remains robust. As $a$ increases, the inter-chain hybridization becomes gradually asymmetric, weakening the horizontal skin accumulation. At the critical point $a=t$, one of the inter-chain hopping amplitudes vanishes, rendering the coupling unidirectional and suppressing the NHSE, thereby allowing disorder-induced Anderson localization to dominate. Upon further increasing $a$, the NHSE re-emerges despite the energy spectrum remaining line-gapped. We therefore uncover successive transitions from skin localization to Anderson localization and back to skin localization by tuning the asymmetric inter-chain coupling. 
\section{Model}\label{Sec2}
We consider a two-leg ladder consisting of Hatano--Nelson chain \cite{PhysRevLett.77.570} coupled to a Hermitian tight-binding chain in presence of disorder \cite{jin2025anderson}. The total Hamiltonian is written as
\begin{align}
H
=&
\sum_{j=1}^{N-1}
\left[
(\gamma+\lambda)
a_j^\dagger a_{j+1}
+
(\gamma-\lambda)
a_{j+1}^\dagger a_j
\right]
\nonumber\\
&
+
J
\sum_{j=1}^{N-1}
\left(
b_j^\dagger b_{j+1}
+
b_{j+1}^\dagger b_j
\right)
\nonumber\\
&
+
\sum_{j=1}^{N}
\left[
(t-a)
a_j^\dagger b_j
+
(t+a)
b_j^\dagger a_j
\right]
\nonumber\\
&
+
\sum_{j=1}^{N}
\Delta_j
\left(
a_j^\dagger a_j
-
b_j^\dagger b_j
\right).
\end{align}

$a_j^\dagger$ and $b_j^\dagger$ are creation operators of Hatano-Nelson and Hermitian chains. $\gamma\pm \lambda$ and $J$ correspond to asymmetric hopping and symmetric hopping for Hatano-Nelson and Hermitian chains respectively.\\
$t\pm a$ and $\Delta_j$ correspond to asymmetric inter chain coupling and on-site disorder on both the chains.\\
In the asymmetric inter chain coupling the forward and backward vertical hopping amplitudes become,
\begin{equation}
t_{\uparrow}=t-a,
\qquad
t_{\downarrow}=t+a.
\end{equation}
Therefore,
$t_{\uparrow}\neq t_{\downarrow}$ for $a>0$. $a$ introduces an additional source of non-Hermiticity independent of the Hatano--Nelson hopping. By tuning the asymmetric inter-chain coupling parameter $a$, the skin localization is progressively suppressed in the upper (Hatano--Nelson) chain, while it becomes increasingly pronounced in the lower (Hermitian) chain. Simultaneously, the variation in the asymmetric inter-chain coupling parameter $a$ gives rise to an Anderson-localized regime at critical point $a=t$.
\section{Results}
\begin{figure*}[t]
\centering
\includegraphics[width=\textwidth]{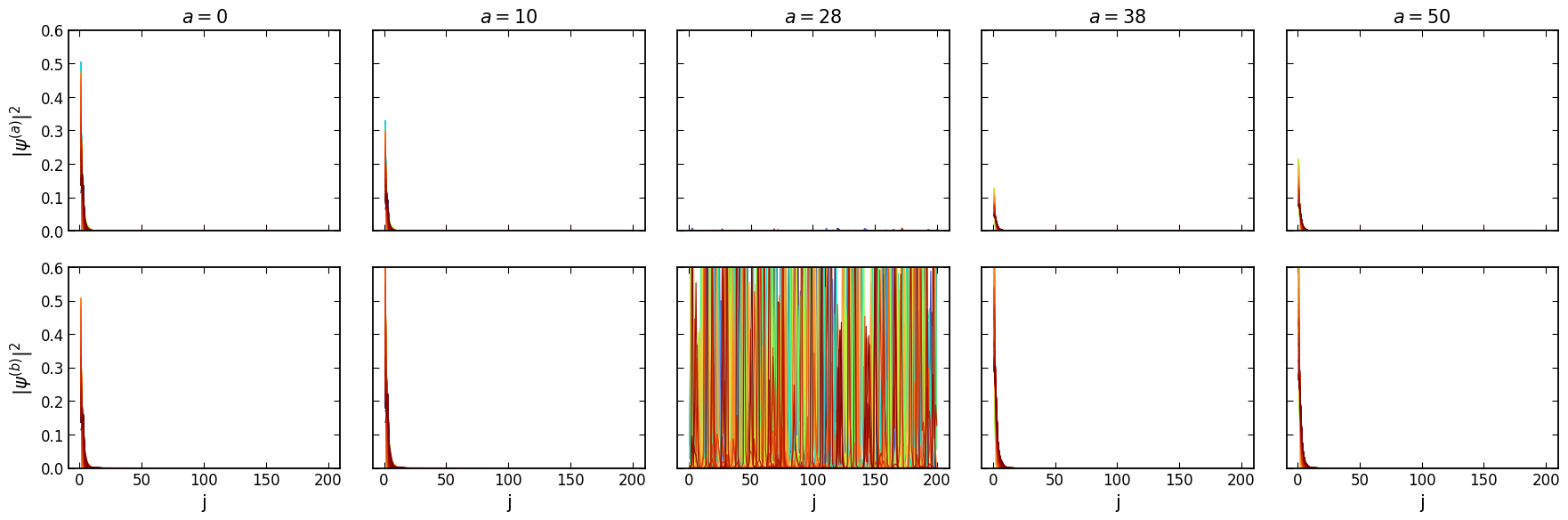}
\caption{
Spatial probability distributions of all right eigenstates in the (top) Hatano--Nelson chain and (bottom) Hermitian chain for $a=0$, $10$, $28$, $38$, and $50$. The eigenstates exhibit boundary localization for $a<t$ and $a>t$, characteristic of the non-Hermitian skin effect. At the critical point $a=t=28$, the suppression of skin localization results in spatially distributed eigenstates associated with disorder-induced Anderson localization. Parameters: $N=200$, $t=28$, $\gamma=1.0$, $\lambda=1.5$, $J=1.0$, and $W=12$.
}
\label{fig:eigenstates}
\end{figure*}
The additional control parameter $a$ in the inter-chain coupling effectively competes with the asymmetric hopping in the non-Hermitian chain. By tuning $a$, the relative strength of these competing non-reciprocal processes can be controlled, leading to significant modifications in both the spectral and localization properties of the system. 

\begin{figure}[H]
\centering
\includegraphics[width=1.0\linewidth]{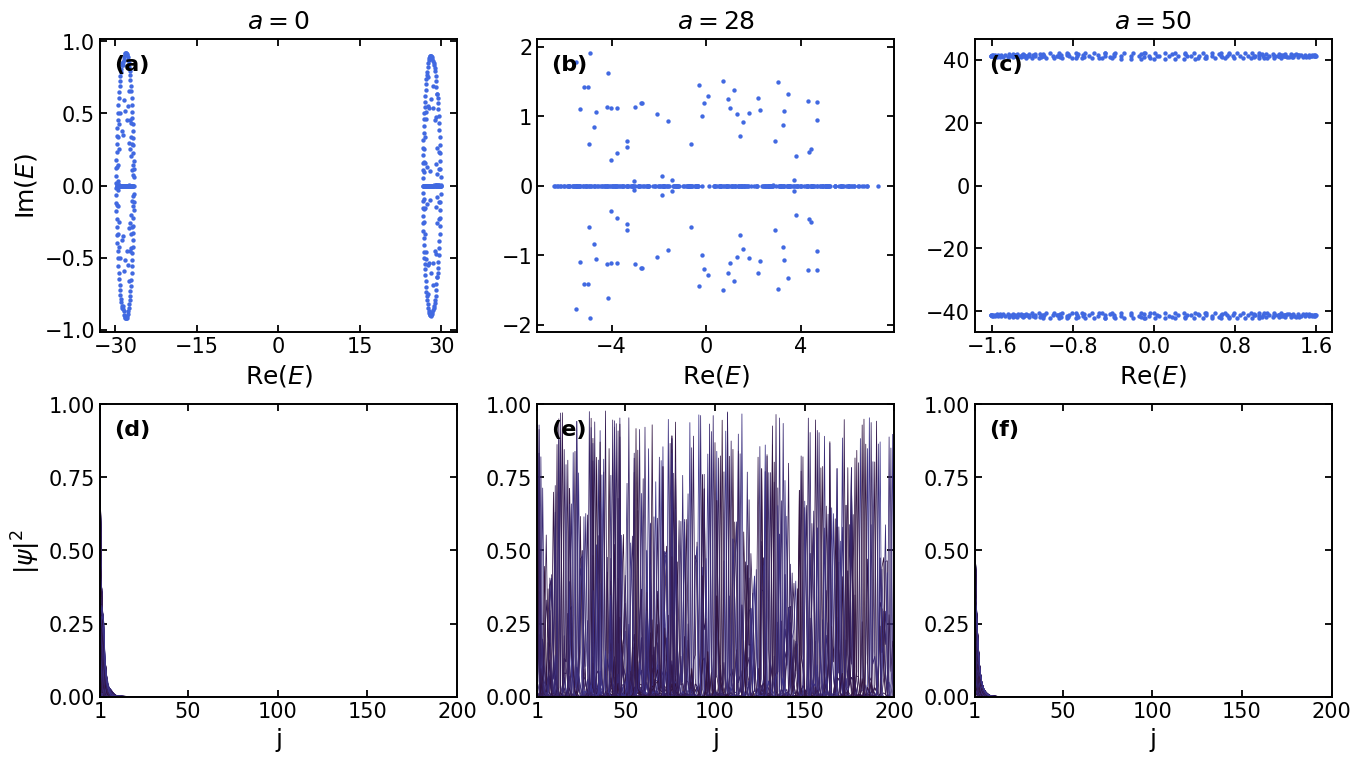}
         \caption{
Complex energy spectra (top row) and corresponding right eigenstate probability distributions of in lower chain (bottom row) for (a,d) $a=0$, (b,e) $a=28$, and (c,f) $a=50$. At $a=0$, the spectrum exhibits a point-gap topology with skin-localized eigenstates. At the critical value $a=28$, the point gap closes. For $a=50$, the spectrum remains line-gapped, while the eigenstates relocalize at the boundary. Parameters: $N=200$, $\gamma=1$, $\lambda=1.5$, $J=1$, $t=28$, and $W=12.0$.
}
        \label{fig1}
\end{figure}
Figure~\ref{fig:eigenstates} shows the spatial probability distributions of all right eigenstates in the Hatano--Nelson (upper) and Hermitian (lower) chains for different values of the asymmetric inter-chain coupling $a$. For $a=0$, both chains exhibit pronounced boundary localization, indicating that the non-Hermitian skin effect induced by the Hatano--Nelson chain is efficiently transferred to the Hermitian chain through the reciprocal inter-chain coupling. As $a$ increases, the localization of the eigenstates in the Hatano--Nelson chain is progressively suppressed, while the localization in the Hermitian chain becomes more pronounced due to the increasing asymmetry of the inter-chain hopping. At the critical point $a=t$, the skin effect is completely suppressed and the eigenstates become distributed throughout the lattice in the lower chain, consistent with the emergence of disorder-induced Anderson localization. Upon further increasing $a$ beyond $t$, the skin effect re-emerges in both chains.\\ 
To gain further insight into the localization properties of the lower chain specifically, we examine the spatial distribution of the eigenstates for different values of the asymmetric inter-chain coupling. As shown in Fig.~\ref{fig1}, in the absence of asymmetric inter-chain coupling ($a=0$), the complex energy spectrum exhibits a point-gap topology enclosing the reference point $E_b$, while the corresponding eigenstates are predominantly localized at the left boundary, consistent with the conventional NHSE. Increasing the asymmetric inter-chain coupling a continuously modifies both the spectral topology and the localization properties. The asymmetric inter chain coupling redistributes the localization weight between the two chains, and at the critical value $t=a$, the point gap collapses, signaling a transition. For $a>t$, the spectrum develops an imaginary line-gap instead of a point gap. Surprisingly, however, the eigenstates remain localized near the boundary, indicating that the NHSE persists even in the line-gap regime. This finding suggests that, in the present disordered non-Hermitian ladder, boundary localization is not solely dictated by point-gap topology and points toward a more general mechanism for the emergence of the NHSE beyond the conventional point-gap paradigm. \\
To investigate the localization properties of the eigenstates in the presence of disorder, we compute the disorder-averaged inverse participation ratio (IPR) and the mean center of mass ($m_{\mathrm{com}}$) using the right eigenvectors of the Hamiltonian.\\
Let $\ket{\psi_n}$ denote the $n^{th}$ normalized right eigenstate of the Hamiltonian,
\begin{equation}
\sum_{j=1}^{N}\left(|\psi_{A,j}^{(n)}|^2+|\psi_{B,j}^{(n)}|^2\right)=1,
\end{equation}
where $\psi_{A,j}^{(n)}$ and $\psi_{B,j}^{(n)}$ represent the amplitudes of the eigenstate on the upper and lower chains, respectively.\\
The inverse participation ratio (IPR) of the $n$th eigenstate is defined as
\begin{equation}
\mathrm{IPR}_n=
\sum_{j=1}^{N}
\left(
|\psi_{A,j}^{(n)}|^4
+
|\psi_{B,j}^{(n)}|^4
\right).
\label{eq:ipr}
\end{equation}
The IPR measures the degree of localization of an eigenstate. For an extended state, the IPR scales approximately as $1/(2N)$ and therefore approaches zero in the thermodynamic limit, whereas it remains finite for a localized state.\\
The average IPR for a given disorder realization is obtained by averaging Eq.~(\ref{eq:ipr}) over all $2N$ eigenstates,
\begin{equation}
\overline{\mathrm{IPR}}
=
\frac{1}{2N}
\sum_{n=1}^{2N}
\mathrm{IPR}_n.
\label{eq:avgipr}
\end{equation}
Finally, the disorder-averaged IPR is obtained by averaging $\overline{\mathrm{IPR}}$ over $N_r$ independent disorder realizations,
\begin{equation}
\langle\mathrm{IPR}\rangle
=
\frac{1}{N_r}
\sum_{r=1}^{N_r}
\overline{\mathrm{IPR}}^{(r)},
\end{equation}
where $N_r$ denotes the total number of disorder realizations.\\
To quantify the average spatial distribution of the eigenstates, we define the mean center of mass. For a given disorder realization, the average probability density at lattice site $j$ is
\begin{equation}
A_j=
\frac{1}{2N}
\sum_{n=1}^{2N}
\left(
|\psi_{A,j}^{(n)}|^2
+
|\psi_{B,j}^{(n)}|^2
\right).
\label{eq:Aj}
\end{equation}
The mean center of mass is then calculated as
\begin{equation}
m_{\mathrm{com}}
=
\frac{\displaystyle\sum_{j=1}^{N}jA_j}
{\displaystyle\sum_{j=1}^{N}A_j}.
\label{eq:mcom}
\end{equation}
Finally, the disorder-averaged mean center of mass is given by
\begin{equation}
\langle m_{\mathrm{com}}\rangle
=
\frac{1}{N_r}
\sum_{r=1}^{N_r}
m_{\mathrm{com}}^{(r)}.
\end{equation}
The quantity $\langle\mathrm{IPR}\rangle$ provides a measure of the average localization strength of the eigenstates, while $\langle m_{\mathrm{com}}\rangle$ characterizes the average position of the probability density along the lattice. Together, these quantities provide complementary information regarding the localization properties and spatial redistribution of the eigenstates as the system parameters are varied.
\begin{figure}[H]
\centering
\includegraphics[width=1.0\linewidth]{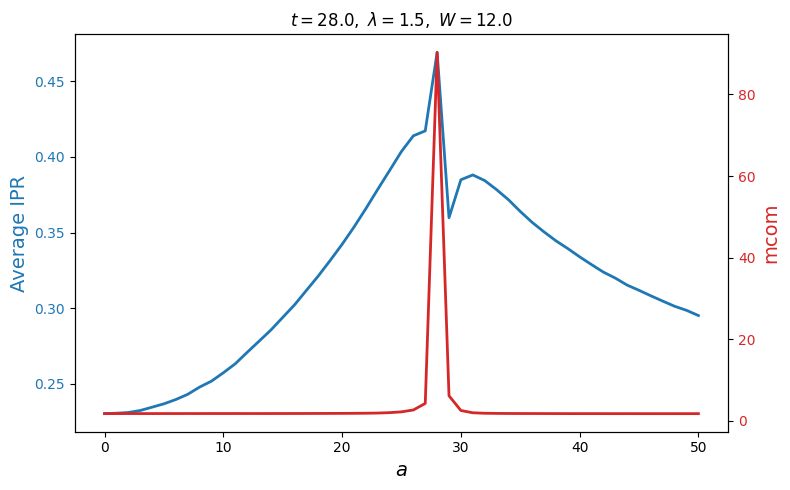}
        \caption{
Disorder-averaged inverse participation ratio (IPR) and mean center of mass ($m_{\mathrm{com}}$) as functions of the asymmetric inter-chain coupling $a$.  Parameters: $N=200$, $\gamma=1$, $\lambda=1.5$, $J=1$, $t=28$, and $W=12$, averaged over 50 disorder realizations.
}
        \label{fig2}
\end{figure}
Figure~\ref{fig2} illustrates the variation of the disorder-averaged IPR and the $m_{\mathrm{com}}$ as functions of the inter-chain coupling parameter $a$. For $a<t$, the average IPR increases steadily with increasing $a$, indicating that the skin modes become progressively less localized and spread over a larger portion of the lattice. As $a$ approaches $t$, the localization reaches its weakest, reflected by the maximum value of the average IPR.\\
 At the critical point $a=t$, the inter-chain coupling becomes unidirectional, which is accompanied by a pronounced peak in $m_{\mathrm{com}}$. This sharp enhancement signifies a sudden redistribution of the eigenstate weight across the lattice and marks the transition between two distinct localization regimes. Beyond the critical point ($a>t$), the average IPR decreases gradually, indicating that the eigenstates recover stronger localization. Simultaneously, $m_{\mathrm{com}}$ rapidly returns to its baseline value, implying that the abrupt shift in the $m_{\mathrm{com}}$ occurs only at the unidirectional coupling point. These results demonstrate that the parameter $a$ provides an effective means of tuning the localization properties of the skin modes, with the most dramatic change occurring at the critical point $t=a$ corresponding to unidirectional inter-chain coupling.\\
 Figure~\ref{fig3} shows the variation of the disorder-averaged IPR as a function of the system size $N$ for different values of the inter-chain coupling parameter $a$. The finite-size scaling of the IPR provides a useful criterion for distinguishing between extended and localized eigenstates. For an extended state, the wave function is distributed over the entire lattice, and consequently the IPR decreases with increasing system size, scaling as
\begin{equation}
\mathrm{IPR}\sim \mathcal{O}(N^{-1}),
\end{equation}
approaching zero in the thermodynamic limit. In contrast, for a localized state, the wave function remains confined to a finite region of the lattice independent of the system size. As a result, the IPR remains finite and scales as
\begin{equation}
\mathrm{IPR}\sim \mathcal{O}(1).
\end{equation}
\begin{figure}[H]
\centering
\includegraphics[width=0.9\linewidth]{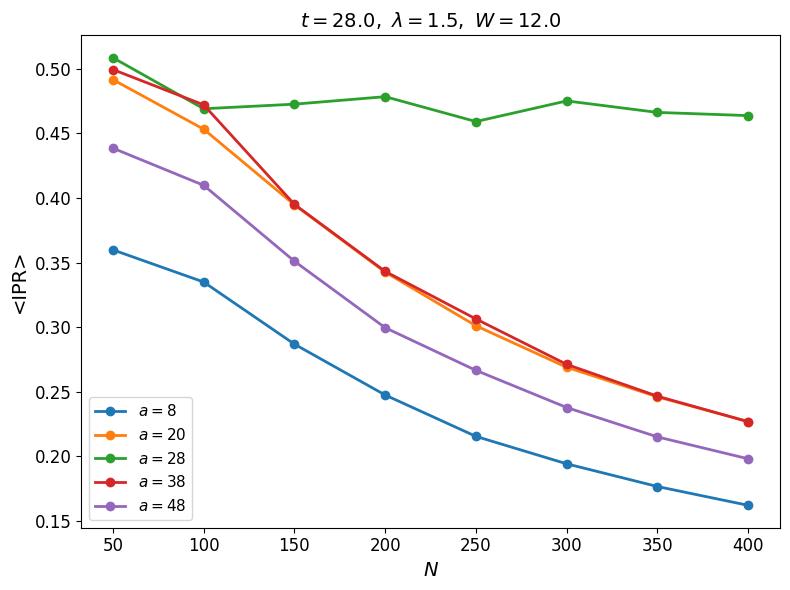}
         \caption{
Disorder-averaged inverse participation ratio (IPR) as a function of the system size $N$ for different values of the asymmetric inter-chain coupling $a$. The nearly constant IPR at $a=28$ indicates localization, while the decreasing IPR for other values of $a$ reflects delocalization. Parameters: $\gamma=1$, $\lambda=1.5$, $J=1$, $t=28$, and $W=12$, averaged over 10 disorder realizations.
}
        \label{fig3}
\end{figure}

As shown in Fig.~\ref{fig3}, for $a=8$, $20$, $38$, and $48$, the average IPR decreases monotonically with increasing system size, indicating that the corresponding eigenstates become progressively more extended. On the other hand, for the critical value $a=28$, the average IPR remains nearly independent of the system size and exhibits only minor fluctuations arising from finite-size and disorder effects. The absence of any size dependence indicates that the eigenstates remain localized even in the thermodynamic limit. These results clearly indicate that the localization properties undergo a qualitative change near the unidirectional coupling point, where the skin modes retain their localized nature despite increasing system size.
\section{Phase diagram}
\begin{figure}[H]
\centering
\includegraphics[width=1.0\linewidth]{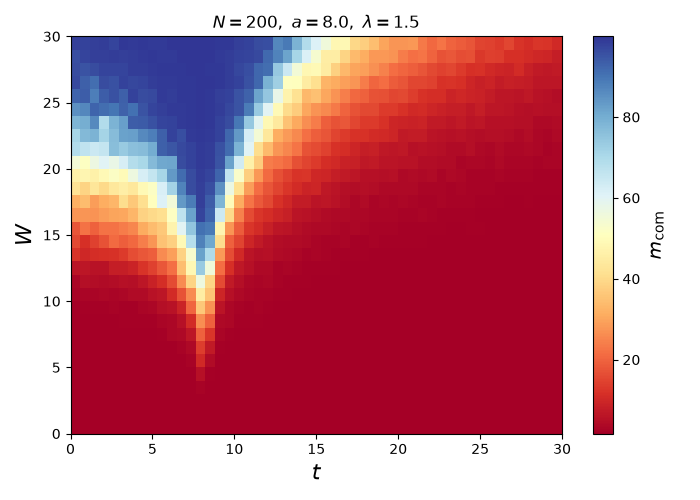}
\caption{Phase diagram of the center-of-mass parameter $m_{\rm com}$ in the $(t,W)$ plane for $a=8$. The V-shaped high-$m_{\rm com}$ region centred at $t=a$ marks the suppression of the NHSE and the emergence of disorder-dominated localization. Parameters: $N=200$, $\gamma=1.0$, $\lambda=1.5$, and $J=1.0$.}
\label{fig02}
\end{figure}

In this section we explore the phase diagram of disorder verses t for a fixed value of a. By this we can understand the transition between skin localized phase to Anderson phase.\\ 
Figure~\ref{fig02} presents the phase diagram of the $m_{\rm com}$ in the $(t,W)$ plane for a fixed asymmetric inter-chain coupling $a=8$. According to the adopted definition, larger values of $m_{\rm com}$ correspond to Anderson localization, while smaller values indicate that the eigenstates are preferentially accumulated near the system boundary due to the NHSE.
A pronounced V-shaped region of enhanced $m_{\rm com}$ develops around the critical condition $t=a$. At this point, one of the inter-chain hopping amplitudes vanishes ($t-a=0$), rendering the inter-chain coupling unidirectional. Consequently, the NHSE is strongly suppressed and disorder becomes the dominant localization mechanism, giving rise to an Anderson-localized regime. As $t$ deviates from $a$, the $m_{\rm com}$ parameter decreases, indicating that the localization centre progressively shifts back towards the system boundary, consistent with the recovery of the NHSE. Moreover, increasing the disorder strength broadens the high-$m_{\rm com}$ region, demonstrating that disorder enlarges the parameter regime over which Anderson localization prevails.
\begin{figure}[H]
\centering
\includegraphics[width=1.0\linewidth]{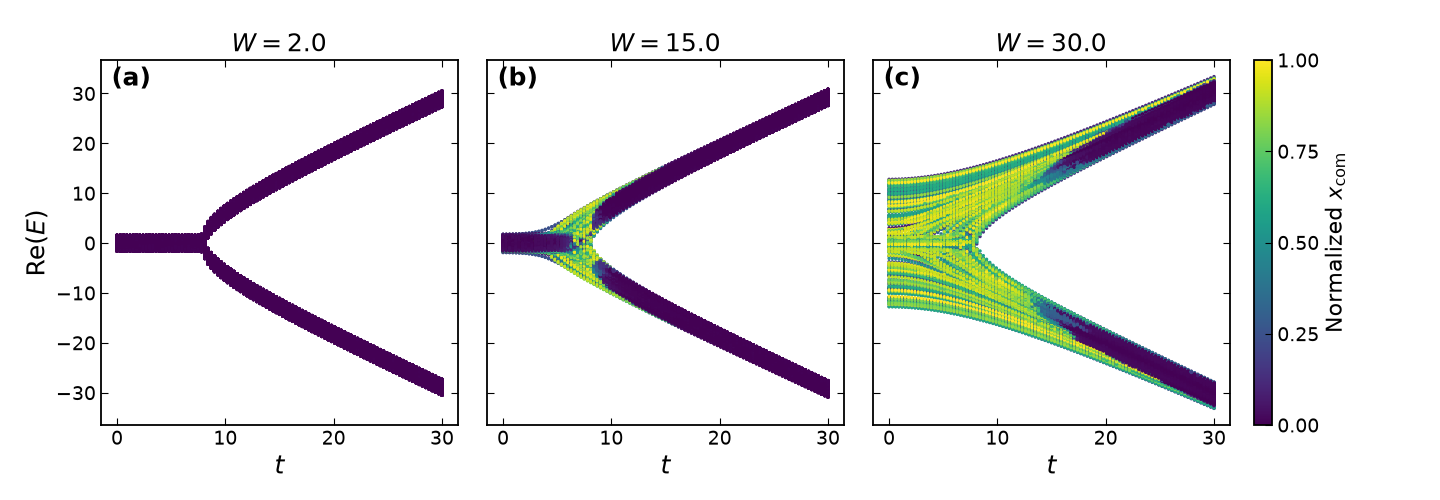}
\caption{Real part of the energy spectrum as a function of $t$ for (a) $W=2$, (b) $W=15$, and (c) $W=30$. Eigenstates are color coded by the normalized center of mass, illustrating the evolution of the localization with increasing disorder. Parameters: $N=200$, $a=8$, $\gamma=1.0$, $\lambda=1.5$, and $J=1.0$.}
\label{fig03}
\end{figure}
To gain further insight into the evolution of the localization properties, Fig.~\ref{fig03} shows the real-energy spectrum as a function of $t$ for three representative disorder strengths, with each eigenstate color coded by its normalized center of mass. For weak disorder [Fig.~\ref{fig03}(a)], the eigenstates remain predominantly localized near the boundary over the entire parameter range, reflecting the robustness of the NHSE. 
\begin{figure}[H]
\centering
\includegraphics[width=0.9\linewidth]{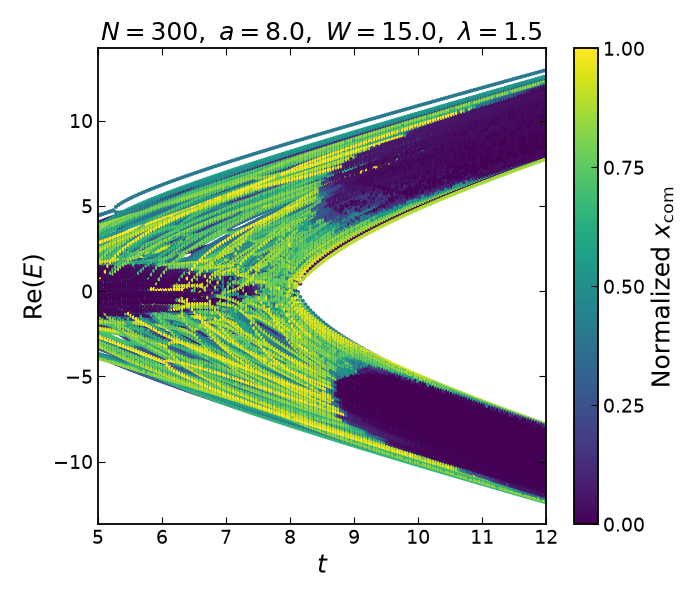}
\caption{Real part of the energy spectrum as a function of $t$ for (a) $W=2$, (b) $W=20$, and (c) $W=30$. Eigenstates are colour coded by the normalized center of mass, illustrating the evolution of the localization centre with increasing disorder. Parameters: $N=200$, $a=8$, $\gamma=1.0$, $\lambda=1.5$, and $J=1.0$.}
\label{fig04}
\end{figure}
As the disorder strength increases [Fig.~\ref{fig03}(b)], eigenstates with $m_{\rm com}$ distributed throughout the bulk gradually emerge in the vicinity of $t=a$, consistent with the increase of $m_{\rm com}$ observed in Fig.~\ref{fig02}. For stronger disorder [Fig.~\ref{fig03}(c)], the bulk-localized states occupy a substantially wider parameter region, indicating that Anderson localization becomes increasingly dominant. Nevertheless, away from the vicinity of $t=a$, eigenstates with boundary-localized centers of mass remain visible, demonstrating the persistence of the NHSE outside the disorder-dominated regime.

The $m_{\rm com}$ of an individual eigenstate is defined as
\begin{equation}
x_{\rm com}=\sum_{j=1}^{N}j\left(|\psi_a(j)|^2+|\psi_b(j)|^2\right),
\end{equation}
where the eigenstate is normalized such that $\sum_j\left(|\psi_a(j)|^2+|\psi_b(j)|^2\right)=1$. To facilitate comparison across different system sizes, we employ the normalized center-of-mass,
\begin{equation}
\tilde{x}_{\rm com}=\frac{x_{\rm com}-1}{N-1},
\end{equation}
which takes values in the interval $[0,1]$. Here, $\tilde{x}_{\rm com}=0$ and $\tilde{x}_{\rm com}=1$ correspond to skin and Anderson localization, respectively. 
\section{Local spectral analysis of the inter-chain coupling}
To gain analytical insight into the role of the asymmetric inter-chain coupling specifically on the local Hamiltonian at every site, we write the local Hamiltonian as,
\begin{equation}
H_j=
\begin{pmatrix}
\Delta_j & t+a\\
t-a & -\Delta_j
\end{pmatrix},
\label{eq:localham}
\end{equation}
where $\Delta_j$ denotes the onsite disorder, $t$ is the reciprocal inter-chain hopping amplitude, and $a$ controls the non-reciprocity of the inter-chain coupling.
The eigenvalues are obtained from the characteristic equation
\begin{equation}
\det(H_j-EI)=0,
\end{equation}
which yields
\begin{equation}
(\Delta_j-E)(-\Delta_j-E)-(t+a)(t-a)=0.
\end{equation}
Using the identity
\begin{equation}
(t+a)(t-a)=t^2-a^2,
\end{equation}
the energy spectrum becomes
\begin{equation}
E_{\pm}
=
\pm
\sqrt{\Delta_j^2+t^2-a^2}.
\label{eq:localspectrum}
\end{equation}
Equation~(\ref{eq:localspectrum}) immediately reveals three distinct spectral regimes.
For
\begin{equation}
a^2<t^2+\Delta_j^2,
\end{equation}
the eigenvalues are purely real,
\begin{equation}
E_{\pm}
=
\pm
\sqrt{\Delta_j^2+t^2-a^2},
\end{equation}
indicating that the local spectrum remains entirely real while the energy gap decreases continuously with increasing asymmetry parameter $a$.
At the critical value
\begin{equation}
a_c=\sqrt{t^2+\Delta_j^2},
\label{eq:critical}
\end{equation}
both eigenvalues merge at zero energy,
\begin{equation}
E_+=E_-=0.
\end{equation}
At this point the Hamiltonian becomes defective and the two eigenvectors coalesce, signaling the emergence of an local exceptional point.
For
\begin{equation}
a^2>t^2+\Delta_j^2,
\end{equation}
the spectrum bifurcates into a complex-conjugate pair,
\begin{equation}
E_{\pm}
=
\pm i
\sqrt{a^2-t^2-\Delta_j^2},
\end{equation}
indicating a transition from a purely real spectrum to a purely imaginary one. 
\begin{figure}[H]
\centering
\includegraphics[width=1.0\linewidth]{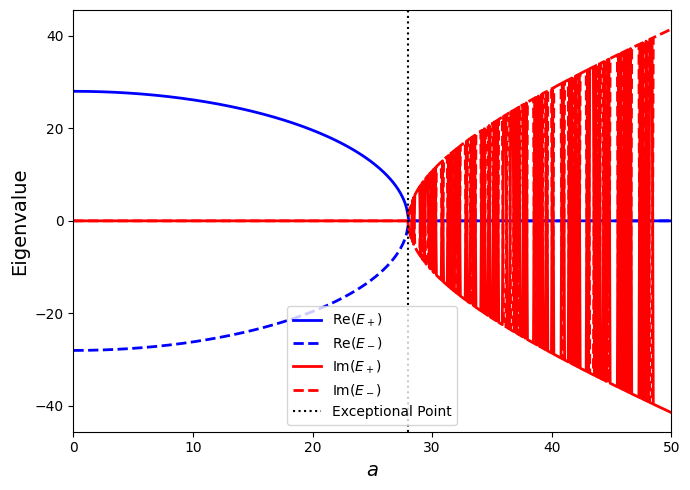}
\caption{Eigenvalue spectrum of the local $2\times2$ Hamiltonian as a function of the asymmetry parameter $a$. The real parts of the eigenvalues (blue) decrease continuously with increasing $a$ and coalesce at the local exceptional point ($a_c=\sqrt{t^2+\Delta_j^2}$).}
\label{fig4}
\end{figure}
Consequently, increasing the inter-chain asymmetry drives the local Hamiltonian through an exceptional point, where the spectral gap closes before reopening along the imaginary-energy axis.
The complete evolution of the local spectrum is illustrated in Fig.~\ref{fig4}. As the asymmetry parameter $a$ increases, the real parts of the eigenvalues gradually approach each other and eventually coalesce at the exceptional point. Beyond this critical value, the real spectrum disappears and the eigenvalues split into purely imaginary branches with opposite signs. This behavior demonstrates that the asymmetric inter-chain coupling alone is sufficient to induce a non-Hermitian phase transition at the local level.
The corresponding right eigenvectors satisfy
\begin{equation}
H_j|\psi_{\pm}^{R}\rangle
=
E_{\pm}|\psi_{\pm}^{R}\rangle,
\end{equation}
and are given by
\begin{equation}
|\psi_{\pm}^{R}\rangle
=
\mathcal{N}_{\pm}
\begin{pmatrix}
1\\
\dfrac{E_{\pm}-\Delta_j}{t+a}
\end{pmatrix},
\label{eq:eigenvector}
\end{equation}
where $\mathcal{N}_{\pm}$ is the normalization constant. At the local exceptional point, shown in Fig.~\ref{fig4}, both eigenvectors merge into a single state, confirming the non-Hermitian degeneracy of the local Hamiltonian. The simultaneous coalescence of both eigenvalues and eigenvectors is the defining characteristic of an exceptional point and distinguishes it from an ordinary Hermitian degeneracy.
Although this analysis neglects the intra-chain hopping amplitudes $(\gamma\pm\lambda)$ and $J$ since $\Delta_j$ and $t$ are dominant in terms of strength, it captures the essential role of the asymmetric inter-chain coupling. Figure~\ref{fig4} clearly demonstrates that increasing $a$ continuously drives the local Hamiltonian through an exceptional point, separating two qualitatively distinct spectral phases. In the complete ladder Hamiltonian, the local two-level systems are coupled through the intra-chain hopping terms to form dispersive bands. Nevertheless, the local analysis provides a transparent analytical understanding of the numerical results presented in the following sections, where increasing the inter-chain asymmetry governs the spectral evolution and strongly influences the localization properties of the full non-Hermitian ladder.
\section{Conclusion}
In this work, we have investigated the interplay between disorder and non-Hermitian effects in a two-leg ladder consisting of a non-Hermitian Hatano--Nelson chain coupled to a Hermitian chain through asymmetric inter-chain hopping. By introducing an additional control parameter in the inter-chain coupling, we demonstrated that the competition between the intrinsic non-reciprocity of the Hatano--Nelson chain and the asymmetric inter-chain coupling gives rise to rich localization behavior in the presence of strong disorder.\\
Our results reveal that tuning the asymmetric inter-chain coupling drives successive transitions in the nature of the eigenstates. For weak asymmetry, the eigenstates exhibit the conventional non-Hermitian skin effect and remain localized at one boundary of the system. At the critical value corresponding to unidirectional inter-chain coupling, the skin localization is destroyed and the eigenstates become Anderson localized. Upon further increasing the asymmetry, the skin effect re-emerges despite the system showing line-gap topology in the spectrum which  demonstrates that robust skin localization can persist in the two-legged ladder system. The finite-size scaling of the inverse participation ratio confirms the localized nature of the eigenstates in this regime, while the behavior of the mean center of mass clearly identifies the transition between the different localization phases.\\
This finding provides new insight into the relationship between spectral topology and localization in non-Hermitian lattices. 
The present work opens several directions for future investigation. It would be interesting to investigate whether Anderson localization coexists with mobility edges in the model Hamiltonian considered.\\ It would also be interesting to explore the role of interactions, quasiperiodic potentials, and correlated disorder in such hybrid Hermitian--non-Hermitian systems. Furthermore, extending the present analysis to higher-dimensional lattices and experimentally realizable photonic, electrical-circuit, or mechanical platforms may provide additional insight into the robustness and controllability of non-Hermitian localization phenomena.
\section{Acknowledgments}
S. R gratefully acknowledges helpful discussions with Dr. Ranjith Kumar R.

\bibliography{ref}
\end{document}